\documentclass[conference]{IEEEtran}
\IEEEoverridecommandlockouts
\usepackage{cite}
\usepackage{amsmath,amssymb,amsfonts}
\usepackage{algorithmic}
\usepackage{graphicx}
\usepackage{textcomp}
\usepackage{comment}
\usepackage{xcolor}
\def\BibTeX{{\rm B\kern-.05em{\sc i\kern-.025em b}\kern-.08em
    T\kern-.1667em\lower.7ex\hbox{E}\kern-.125emX}}
\begin{document}

\title{Dynamic Coherence-Based EM Ray Tracing Simulations in Vehicular Environments\\

}

\author{\IEEEauthorblockN{Ruichen Wang}
\IEEEauthorblockA{\textit{ECE department} \\
\textit{University of Maryland, College Park}\\
Maryland, USA \\
rwang92@umd.edu}
\and
\IEEEauthorblockN{Dinesh Manocha}
\IEEEauthorblockA{\textit{CS and ECE department} \\
\textit{University of Maryland, College Park}\\
Maryland, USA \\
dmanocha@umd.edu}
}

\maketitle

\begin{abstract}
5G applications have become increasingly popular in recent years as the spread of fifth-generation (5G) network deployment has grown. For vehicular networks, mmWave band signals have been well studied and used for communication and sensing. In this work, we propose a new dynamic ray tracing algorithm that exploits spatial and temporal coherence.We evaluate the performance by comparing the results on typical vehicular communication scenarios with $\bf GEMV^{2}$, which uses a combination of deterministic and stochastic models, and $\bf WinProp$, which utilizes the deterministic model for simulations with given environment information. We also compare the performance of our algorithm on complex, urban models and observe a reduction in computation time by 36\% compared to $\bf GEMV^{2}$ and by 30\% compared to WinProp, while maintaining similar prediction accuracy. 
\end{abstract}


\section{Introduction}
With the rise of millimeter-wave (mmWave) in 5G, many related technologies such as ray tracing (RT) have been studied and updated from sub 6 GHz band to mmWave band. RT can be used in many applications such as field prediction for radio network planning, RT-assist channel estimation for optimal beamforming techniques for mobile backhauling, real-time RT for assisted beamforming, and RT assisted localization \cite{b1}. In addition, mmWave has been studied through measurement campaigns for vehicular communications and sensing and has shown promising potential at high frequencies \cite{b2}\cite{b3}. According to \cite{b4}, the RADar-based COMmunication  (RADCOM) technique enables simultaneous communication and sensing, which may further contribute to other technologies. While RADCOM is generating more research interest, ray tracing has been well studied since the 1990s. However, dynamic ray tracing is a recently developed idea for performing deterministic ray-based prediction in dynamic environments, where the environment shifts or objects in the scenes such as vehicles move. Efficient techniques have been proposed for fast ray tracing in such dynamic environments in sound \cite{b5}\cite{b6} and visual rendering \cite{b7}\cite{b8}. Dynamic ray tracing is also gaining more attention with the dramatic increase of mobile devices/users and the advent of wireless systems of traffic control and safety enforcement in the 5G era \cite{b9}, where the transmitter stays more or less static while the surroundings change. On the other hand, we can also choose to apply ray-based propagation models to vehicular propagation with moving terminals and scatterers, as described in \cite{b10}.\\
In this paper, we present a novel algorithm, which utilizes temporal and spatial coherence, for dynamic ray tracing for EM simulation in large environments and compare its performance in a typical vehicular environment with results from $GEMV^2$ \cite{b21} and $WinProp$ \cite{b17}. Our algorithm uses bounding volume hierarchy (BVH), a tree structure on a set of geometric objects, where all geometric objects that form the leaf nodes of the tree, are wrapped in bounding volumes; it can support efficient operations on sets of geometric objects in ray tracing. We implemented BVH with efficient techniques to recompute or update these hierarchies during each frame \cite{b11} and frame-to-frame coherence along with combining path tracing and radiosity methods \cite{b12}. Section II discusses related works and briefly introduces selected ray tracing software. Section III presents our proposed Dynamic Coherence-based EM ray tracing simulator (DCEM) details and the simulated environment. Section IV compares and discusses the simulation results. Section V concludes the paper and argues that our novel dynamic ray tracing approach is comparable to both academic and industrial ray tracing simulators. It also mentions potential future dynamic and real time ray tracing research. The key contributions of this paper are:
\begin{itemize}
\item We propose the first dynamic coherence-based ray tracing algorithm at EM bands for fast and reliable RT simulations.
\item We develop a new ray tracing software, DCEM, which can work at mmWave frequencies, and show that it speeds up computation time by at least 30\% with comparable prediction accuracy to other simulators like $GEMV^2$ and $WinProp$. 
\item We evaluate the performance of our ray tracing algorithm and system in large vehicular communication environments and highlight the benefits.
\end{itemize}

\section{Related work and available ray tracing software}
Analysis of potential mmWave solutions to 5G has been conducted worldwide, and many measurements for indoor and outdoor scenes are described in \cite{b13} \cite{b14} \cite{b15}. However, the models built based on site-specific data could suffer significantly when moving to another site \cite{b16}. Another approach to building reliable models is an analytical approach, which can achieve high prediction accuracy when the environment is precisely built and computed with refined simulation setups. In our DCEM ray tracing software development process, we evaluate our results with $GEMV^2$ and $WinProp$ since they are 1) widely used for different scenarios and 2) not very difficult to install, with relatively detailed documents for their theory bases/user manual.\\
There are other RT software solutions currently in both industry and academia. We provide a brief description of the selected ray tracing software below and summarize the frequency range and applicable scenario in Table I.\\
\textit{WinProp} \cite{b17} is an RT software from Altair that can support standard RT, Intelligent Ray Tracing (IRT), and Dominant Path models (DPM) at frequencies up to 100 GHz. It is also capable of simulating the spatial variability of the objects in various propagation scenarios.\\
\textit{Wireless Insite} \cite{b18}, developed by Remcom, can provide
ray tracing coupled with empirical and deterministic models for frequencies up to 100 GHz. It is accelerated by dimension reduction algorithms as well as performing accelerated computations using a graphics processing unit (GPU) and multi-threaded central processing unit (CPU) hardware acceleration.\\
\textit{NYUSIM} \cite{b19} is an open-source mmWave channel model simulator developed by NYU Wireless based on years of measurements conducted at various frequencies. It is a generic statistical model that runs fast but does not take in site-specific information. \\
\textit{CloudRT} \cite{b20} is an academic ray tracing tool created by members of the State Key Laboratory of Rail Traffic Control and Safety at Beijing Jiaotong University. This method is aimed at designing a high-performance cloud-based RT simulation platform that supports frequencies up to 325 GHz, includes all kinds of propagation mechanisms and mobile scattering objects, and is accelerated with a space partitioning algorithm and multi-thread computing. \\
\textit{$GEMV^2$} is another academic ray tracing tool developed in \cite{b21} to analyze vehicle-to-vehicle channels in large environments \cite{b22}. This tool can simulate city-wide networks with tens of
thousands of vehicles on commodity hardware, providing hybrid RT along with empirical models. However, this work is based on MATLAB and only evaluated by measurements at 5.9 GHz.
\begin{table}[htbp]
\caption{Capacities comparisons of DCEM and other simulators}
\begin{center}
\begin{tabular}{|c|c|c|}
\hline
\cline{2-3} 
\textbf{Simulator} & \textbf{\textit{Frequency range}}& \textbf{\textit{Environment}} \\
\hline
WinProp& up to 100 GHz& Outdoor/indoor \\
\hline
Wireless Insite& up to 100 GHz& Outdoor/indoor\\
\hline
NYUSIM& up to 100 GHz& Outdoor/indoor \\
\hline
CloudRT& up to 325 GHz& Outdoor \\
\hline
$GEMV^2$& 5.9 GHz& Outdoor \\
\hline
DCEM& up to 72 GHz& Outdoor/indoor \\
\hline
\end{tabular}
\label{tab1}
\end{center}
\end{table}

\section{DCEM and Simulation Scenario}
Before going into the details of RT algorithms, we first define the concept of a ray in EM simulations. A ray is a high frequency approximation of Maxwell's equations for propagating electromagnetic (EM) waves based on the electric and magnetic fields expressions as follows:
\begin{align}
    \Vec{E}(\Vec{r})&=\Vec{e}(\Vec{r})e^{-j\beta_0S(\Vec{r})}\\
    \Vec{H}(\Vec{r})&=\Vec{h}(\Vec{r})e^{-j\beta_0S(\Vec{r})}
\end{align}
where $\Vec{e}(\Vec{r})$ and $ \Vec{h}(\Vec{r})$ are magnitude vectors and $S(\Vec{r})$ is the optical path length or eikonal. When $\beta_0 ->\infty$ and considering a series of wavefronts, the power flow lines perpendicular to the wavefronts are the rays, and they do not intersect if there is no focus point. Thus, the ray trajectory will be a straight line in a homogenous medium. The detailed mathematical derivation steps can be found in \cite{b23}.
In conclusion, the ray, which helps analyze the different propagation mechanisms, is a straight line in a homogenous medium that carries energy and obeys the laws of reflection, transmission, and diffraction. In DCEM, we considered four kinds of rays:
\begin{itemize}
\item Direct rays: If a ray goes from the source to the field point directly, the line of sight (LoS) propagation mechanism will be applied, and the pathloss will be simulated by 

\begin{multline}
   PL^{CI}(f,d)[dB]=FSPL(f,d=1m)[dB]\\+10log_{10}(d)[dB]+AT[dB],    
\end{multline}

where $f$ denotes the carrier frequency in GHz, $d$ is the 3D T-R separation distance, $n$ represents the path loss exponent (PLE), and $AT$ is the attenuation term induced by the atmosphere \cite{b15}.
\item Reflected and transmitted rays: If a ray is reflected or transmitted one or more times before reaching the field point, the ray will be segmented into different parts by reflected or transmitted points and be applied with the directed path loss model for each segment. The study in \cite{b1} shows that at the mmWave band, the signal power will be negligible at a high order of reflections and transmissions. In our proposed RT system, the limit of the maximum number of reflections/transmissions is set to 10 to balance the running time and accuracy by simulations trials.
\item Diffracted rays: The diffracted rays are more complicated than the two types of rays mentioned above, since one incident ray at the geometry edge can lead to a cone of diffracted rays. The uniform theory of diffraction (UTD) \cite{b24} is applied to calculate the diffraction coefficients here, and the detailed steps are followed as described in \cite{b25}. The upper bound of diffraction time simulated is set to 3 in this tool by practice.
\item Diffused and scattered rays: The last type of ray comes from rough surfaces such as building exteriors. The scattered rays will be divided
into specular and non-specular components and simulated by the effective roughness model suggested in \cite{b23} \cite{b26}. 
\end{itemize}
In addition to ray simulation, we also performed phase calculation.\\
For the acceleration methods, DCEM utilized three main algorithms: backward ray tracing from the receiver in the visibility determination step, the use of BVH while minimizing its rebuilding cost between frames, and propagation path caching for better frame coherence \cite{b12}. The idea of frame-to-frame coherence arises from the fact that if the object or transmitter does not drastically change its position between two adjacent frames, then most ray paths found in the previous frame can be reused in the current frame and this formulation can significantly reduce the computation times in low-speed dynamic simulations. The notion of coherence in ray tracing has been used for visual and aural rendering \cite{b27}\cite{b28}, but is not widely known in the context of EM ray tracing. 
\begin{itemize}
\item \textit{Backward ray tracing:} In the backward RT algorithm, rays are cast from the listener or the receiver, rather than from each source. We observe that the reflected/diffracted rays coming from geometric primitives in the vicinity of the receiver, contribute more to the total received power. In fact, when casting rays from the source, only a few may reach the receiver and may not be the most perceptually important paths, resulting in more rays needing to be cast to get necessary propagation paths. The backward RT method can compute all the important paths while shooting fewer rays. The backward RT strategy also benefits from the fact that the number of rays no longer scales linearly with the number of sources. \\Our overall algorithm proceeds as follows: 1) cast a random sphere of sampled rays, 2) record reflections and keep a hash table of visited propagation paths for each depth of reflection (up to a user defined threshold, 10 in DCEM), 3) generate a series of image receiver positions when encountering a new triangle series, 4) check the source to see if there is a valid path to the receiver, 5) if any edge is marked as a diffraction edge, consider the sources that lie in the diffraction shadow region from the receiver's perspective and use the UTD diffraction formulation to determine the point on the edge at which diffraction occurs, then perform path validation back to the receiver as with reflection paths, 6) for each valid propagation path, the system calculates the total distance along the path, the direction of the path from the receiver,
and the total attenuation and phase distortion along the path.
\item \textit{Efficient BVH updating:} BVHs have been widely used to accelerate the performance of ray tracing algorithms \cite{b29}, and we take one step further to efficiently recompute or update these hierarchies during each frame. We do this because rebuilding BVHs is expensive in practice, and we minimize the cost by measuring BVH quality degradation between successive frames. The advantages of our method are: 1) it will update/rebuild at a time when necessary without any scene specific settings, 2) when there is little to no degradation, the rebuild would not be initiated, and 3) it is possible to just rebuild subtrees in some cases. These advantages improve the system computation efficiency; detailed evaluations can be found in \cite{b11}.
\item \textit{Propagation path caching:} The visibility hash tables as
persistent caches are used to accelerate the path finding process from frame to frame. Once the valid paths are found, they are kept and updated until removed, i.e., at the beginning of each frame simulation, all triangle sequences in the hash tables are checked to see if the previous paths are still valid since the positions of source and receiver do not change much between frames. This method has significantly lowered the number of visibility rays cast each frame, leading to a higher overall frame rate and lower latency for real time applications. More details are described in \cite{b30}\cite{b31}.
\end{itemize}
In terms of the input, DCEM can take the environment description files including an ``.obj" file for a description of the geometric primitives in the scene and an ``.mtl" file for the material information of the objects in the environment. The following table highlights the material properties used in DCEM in terms of generating results and comparing with other methods \cite{b17}\cite{b32}. For the outdoor scenario, we do not perform any penetration computations. The transmitter height is set to 2 meters in both cases, the frequency is set to 30 GHz, and the transmit power is 5W for outdoor and 0.5W for indoor scenes. The antenna is assumed to be ideal and omnidirectional. We use the modeled small European town built by Turbosquid \cite{b33} to perform RT simulations.

\begin{table}[htbp]
\caption{Material properties used in out system DCEM}
\resizebox{\columnwidth}{!}{\begin{tabular}{|l|l|l|l|l|}
\hline
\textbf{Material} & \textbf{\textit{Thickness(mm)}}& \textbf{\textit{Reflection coefficient}}& \textbf{\textit{Penetration loss(dB)}} \\
\hline
Wall (Outdoor) & /& 0.8& / \\
\hline
Wall (Indoor) & 10& 0.7& 20 \\
\hline
Glass (Indoor) & 5& 0.74& 5 \\
\hline
\end{tabular}
\label{tab2}
}
\end{table}

\begin{figure}[htbp]
\centerline{\includegraphics[scale=1.1]{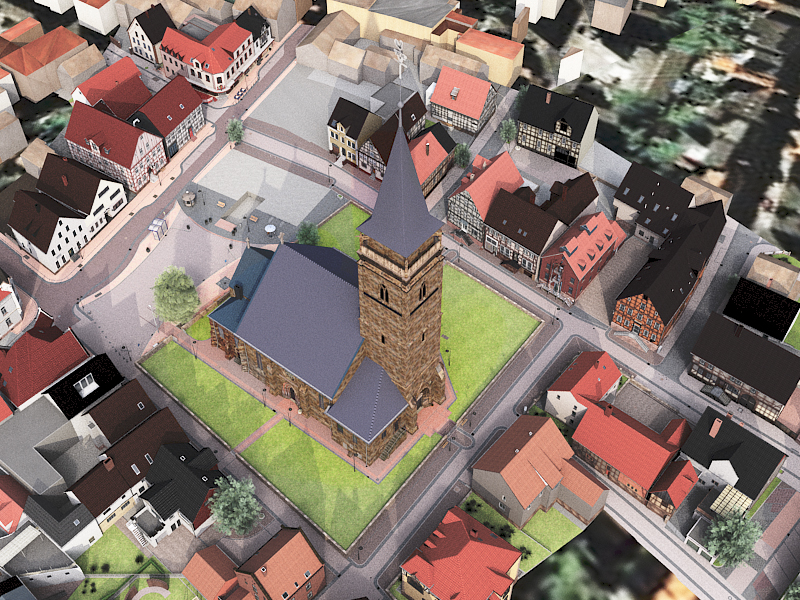}}
\caption{3D model of a European urban model. We show that our method works better than other simulators in such urban environment, with high accuracy and fast computation time.}
\label{fig1}
\end{figure}

\section{Results Comparison and Discussion}
{We highlight some comparison results in this section. The $GEMV^2$ outputs the received signal powers (RSP) of Vehicle-to-everything (V2X) and WinProp can generate heatmaps of RSP. We compare the results with $GEMV^2$ for outdoor vehicular communication scenarios and with WinProp for indoor scenes to evaluate the accuracy of our system DCEM, based on measurements data \cite{b17}\cite{b21}.

\subsection{Outdoor Environment}\label{AA}

We highlight the results and visualizations from $GEMV^2$ in Fig.~\ref{fig2} . The RSP of $GEMV^2$ is a combination of large-scale variations computed by deterministic models and small-scale variations estimated from stochastic models. 
\begin{figure}[htbp]
\centerline{\includegraphics[scale=0.3]{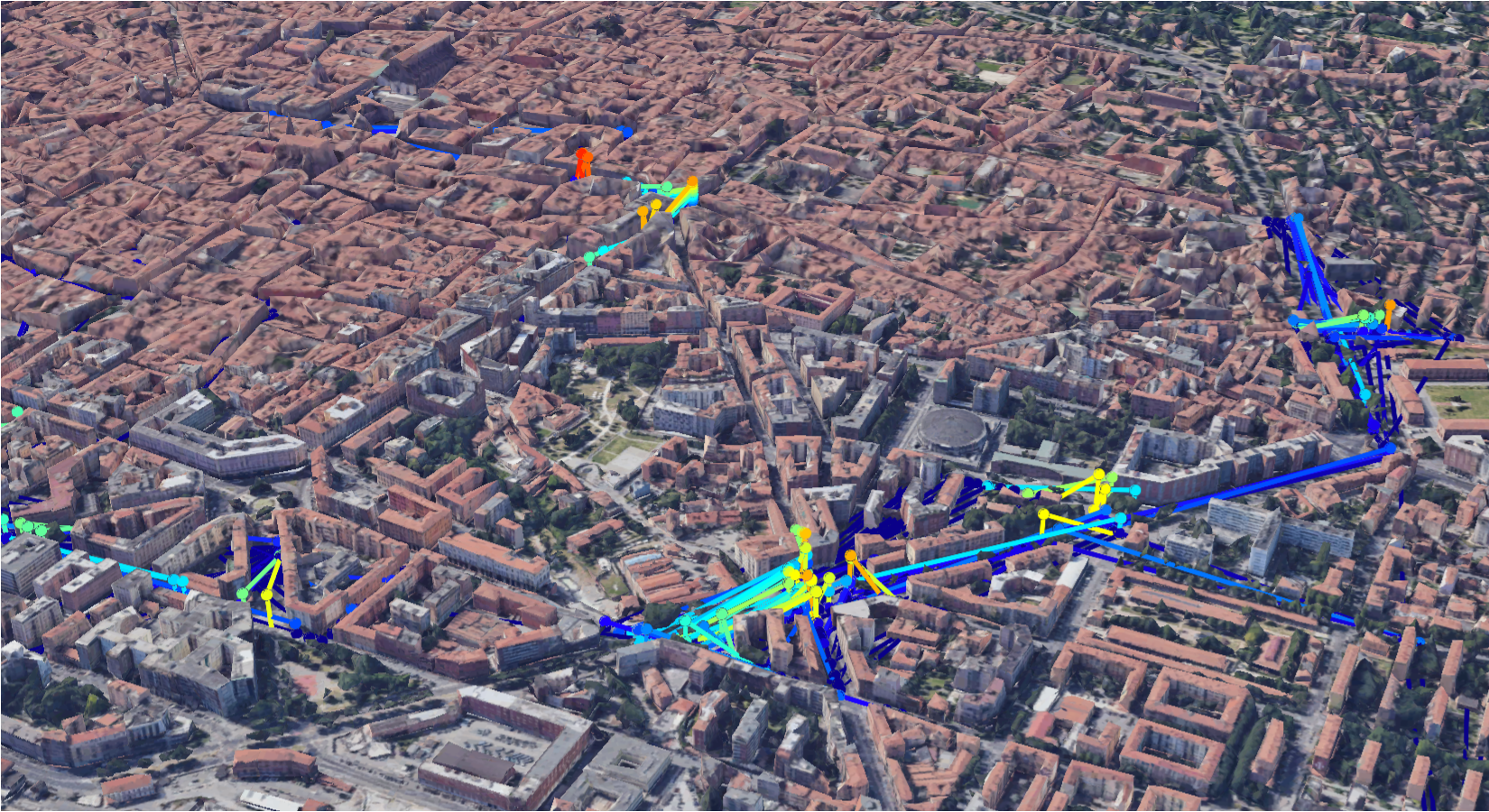}}
\caption{$GEMV^2$ generates V2X simulation output in Keyhole Markup Language (KML) format. We highlight the simulation results of V2V with Google Earth, in the urban Europe model.}
\label{fig2}
\end{figure}

Fig.~\ref{fig3} hightlights a localized region in the urban environment in $GEMV^2$.
\begin{figure}[htbp]
\centerline{\includegraphics[scale=0.27]{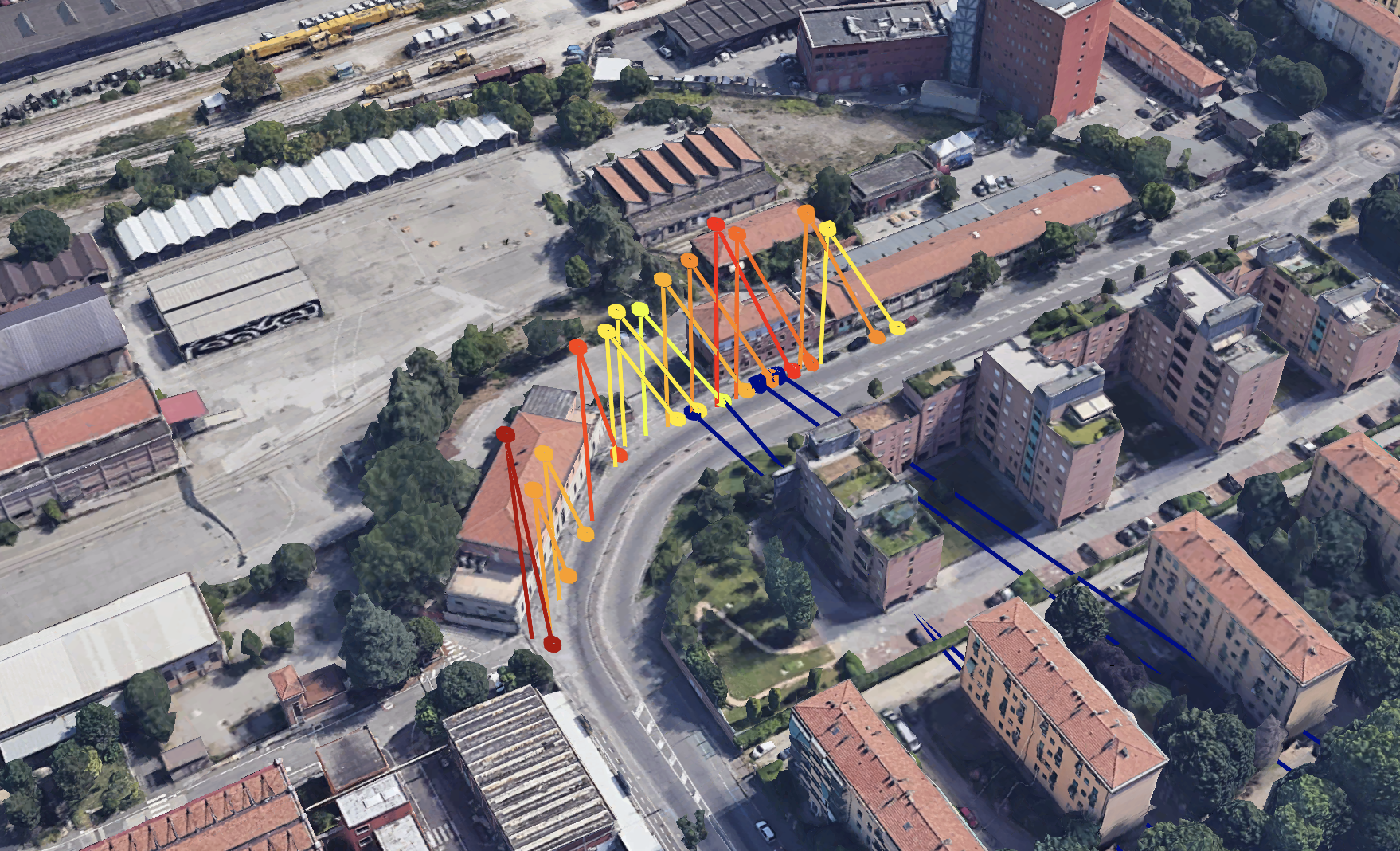}}
\caption{$GEMV^2$ RSP results in V2V, the color and height indicate the RSP stength in this area.}
\label{fig3}
\end{figure}

Fig.~\ref{fig4} shows a simulation comparison between DCEM and $GEMV^2$ within a street in the 3D model urban environment. We define the units and positions of interest to perform simulations in $GEMV^2$ and run the simulation three times to evaluate the small-scale scholastic variations and compare them with the ray tracing results from DCEM. We observe a good match in the predicted positions in RSP and time delay, and the mean of RSP shows an average difference of 1.7 dBm.
\begin{figure}[htbp]
\centerline{\includegraphics[scale=0.11]{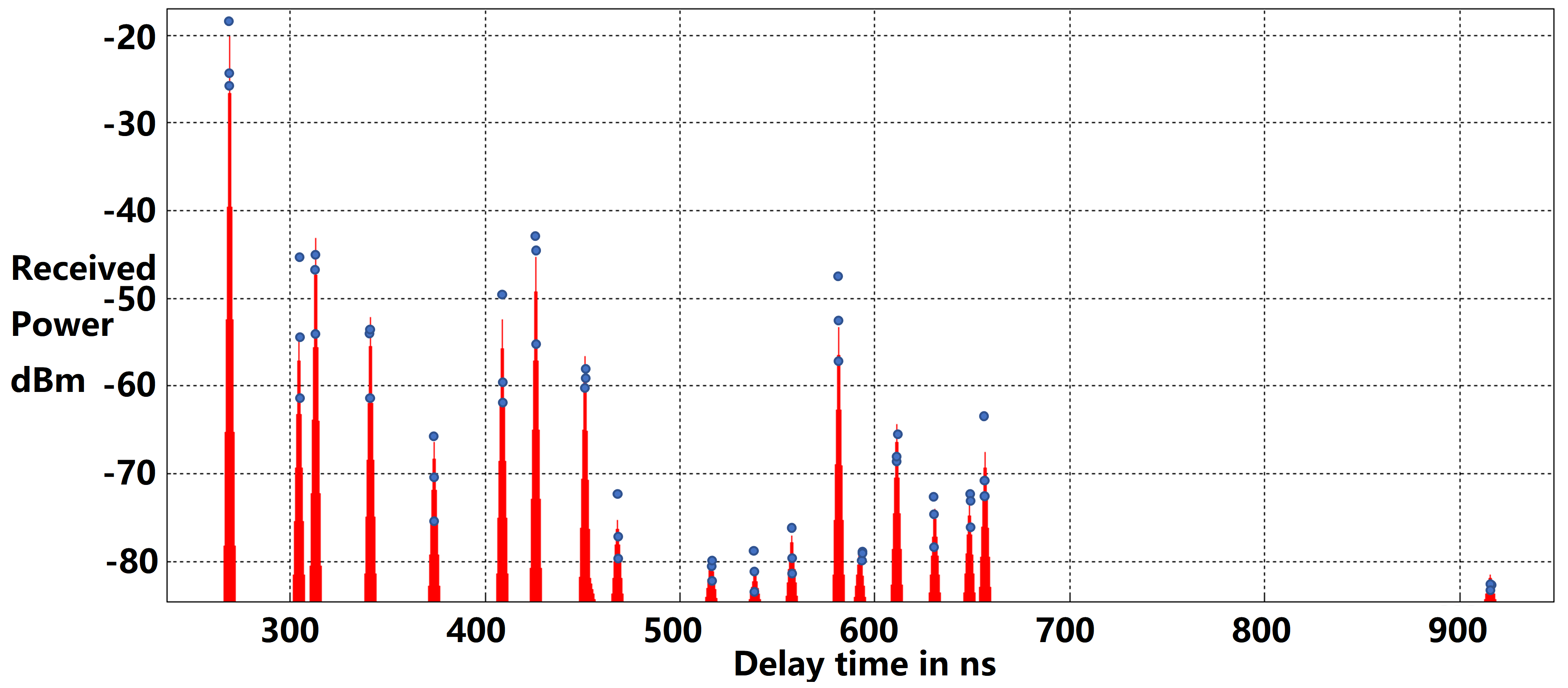}}
\caption{DCEM and $GEMV^2$ RSP results in V2V comparison, the red peaks from DCEM and the blue points from $GEMV^2$. Since the $GEMV^2$ has small scale variation, the simulated result shows a slight variation around the deterministic predictions.}
\label{fig4}
\end{figure}

\subsection{Indoor Environment}\label{AA}
We compare the simulation results of the indoor environments (shown in Fig.~\ref{fig5}) computed using WinProp and DCEM.  The signal frequency is set to 30GHz and the TX height is set at 2 meters. The indoor environment consists of walls, doors, and windows with different reflection/transmission/diffraction coefficients.
\begin{figure}[htbp]
\centerline{\includegraphics[scale=0.7]{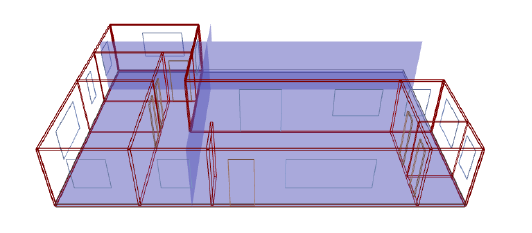}}
\caption{Simple indoor office environment: we compare the performances of WinProp and DCEM indoor because WinProp took a very long time to load and convert the modeled village environment shown in Fig.~\ref{fig1}. This indoor environment with known material information is easy to use for prediction accuracy and performance validation.}
\label{fig5}
\end{figure}

The output one-shot heatmaps of the indoor environment produced by WinProp and DCEM are shown in Fig.~\ref{fig6} and Fig.~\ref{fig7}. We see from the two heatmaps that DCEM predictions show behaviors similar to WinProp results. There is, however, some blurred prediction area towards the boundary of the environment, which might be resulting from different boundary condition settings.

\begin{figure}[htbp]
\centerline{\includegraphics[scale=0.23]{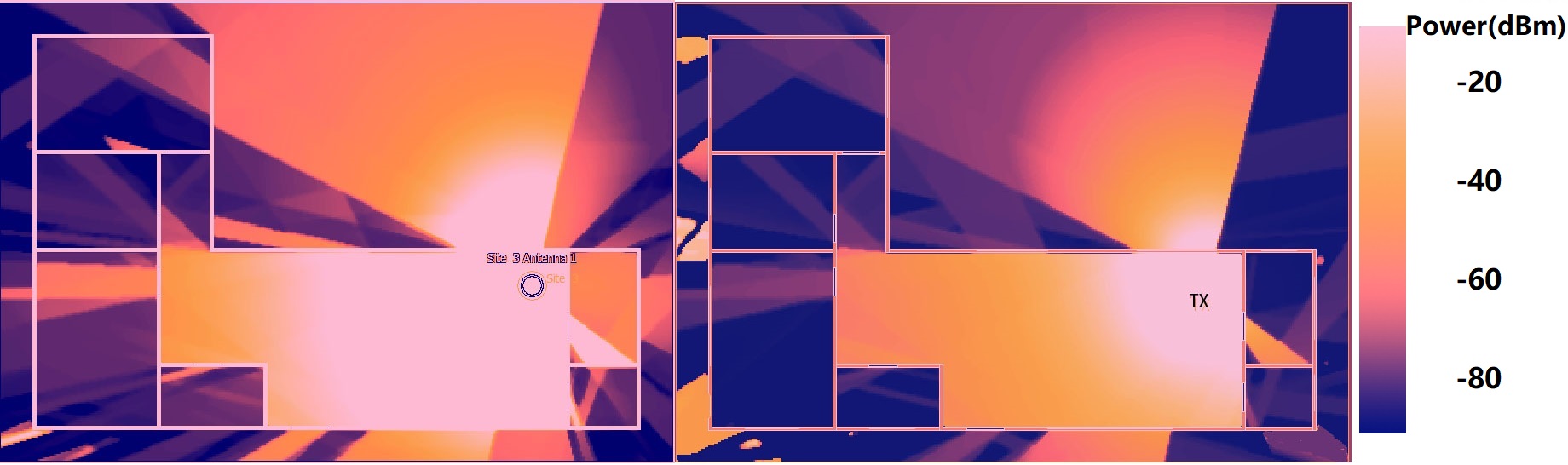}}
\caption{(a) Received power heatmap computed by WinProp (left) using the COST351 model at 30GHz for RT simulation based on balancing the accuracy and running time. The color bar on the right indicates the the strength of received powers. We observe that whew there are windows or doors in the office, some parts of the rays are limited into a cone shape propagation. The material information of walls/doors/windows is provided by WinProp. (b) Received power heatmap by DCEM (right), using our model for predictions at 30 GHz. The same environment and material information is encoded. We can clearly see the reflection at walls, diffraction at edges, and transmission through windows/walls. Due the differences in two models, we observe some differences between the two heatmaps, while the ray behaviors are similar. The accuracy comparison is shown by prediction difference histogram in Fig.~\ref{fig9}.}
\label{fig6}
\end{figure}

A total of three testing cases were simulated with different Tx locations. The other two pairs of heatmap comparisons are shown below with different TX locations in the environment (see Fig.~\ref{fig7} and Fig.~\ref{fig8}).\\
\begin{figure}[htbp]
\centerline{\includegraphics[scale=0.23]{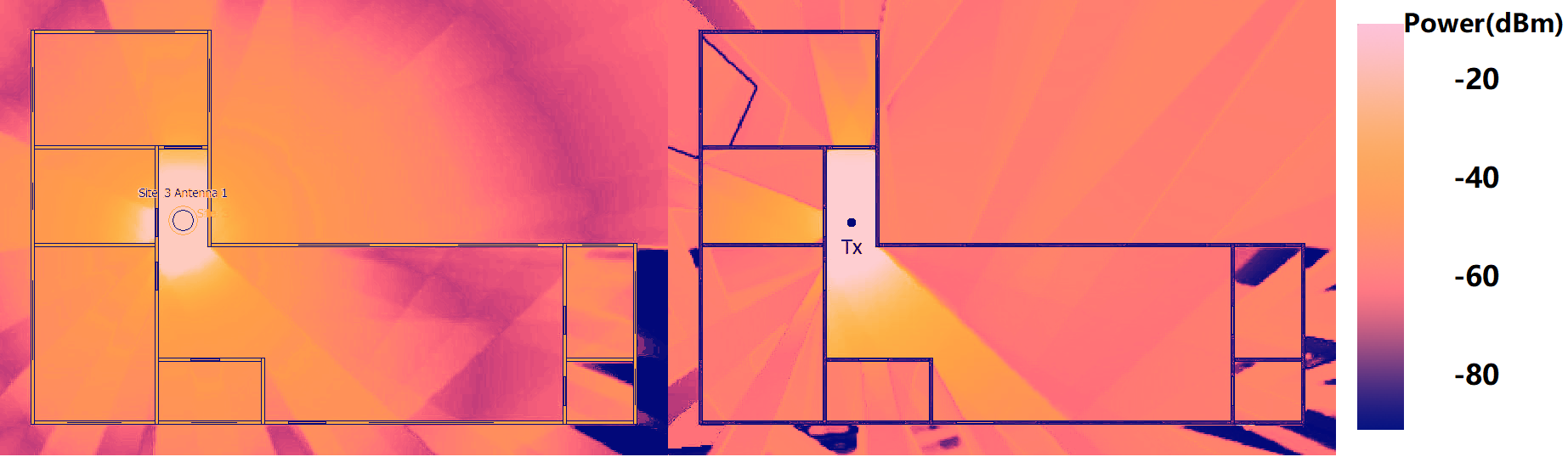}}
\caption{Site 2, received power heatmaps by WinProp (left) and DCEM (right) for the indoor environment with different TX locations. This shows that our method is valid for this indoor environment with accuracy comparable to WinProp.}
\label{fig7}
\end{figure}

\begin{figure}[htbp]
\centerline{\includegraphics[scale=0.23]{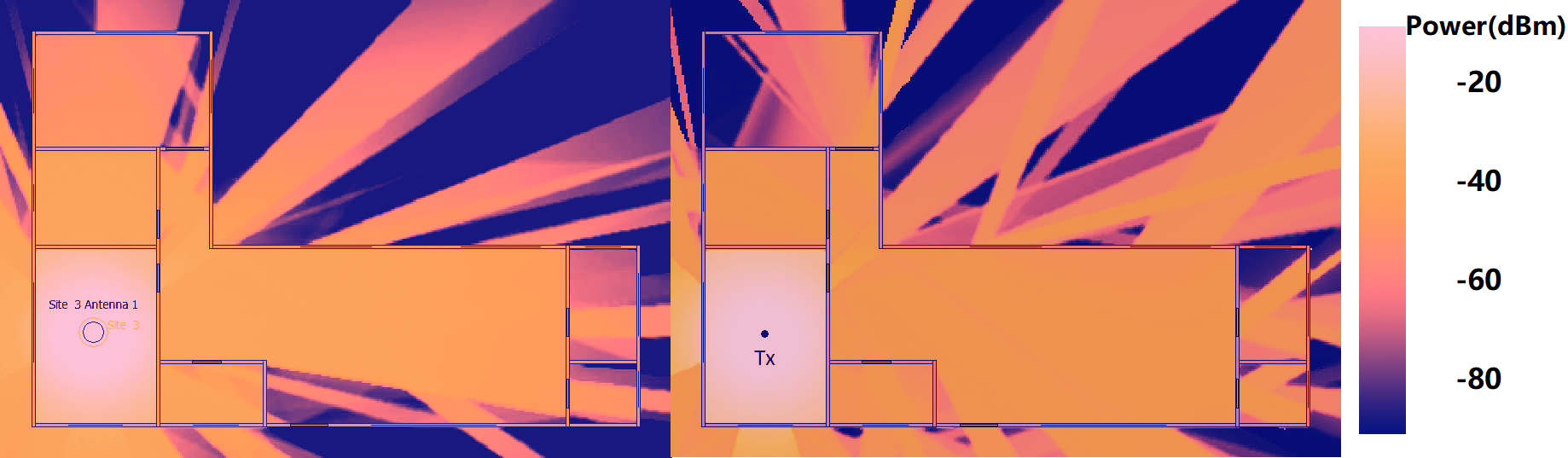}}
\caption{Site 3, received power heatmaps by WinProp (left) and DCEM (right) for the indoor environment with different TX locations.}
\label{fig8}
\end{figure}

The histogram of the difference between the predicted heatmaps from WinProp and DCEM of Site 2 is shown below. According to the numbers, about 58.1\% of the prediction difference is less than 5\% and 95\% of the prediction difference is less than 23.5\%.
\begin{figure}[htbp]
\centerline{\includegraphics[scale=0.3]{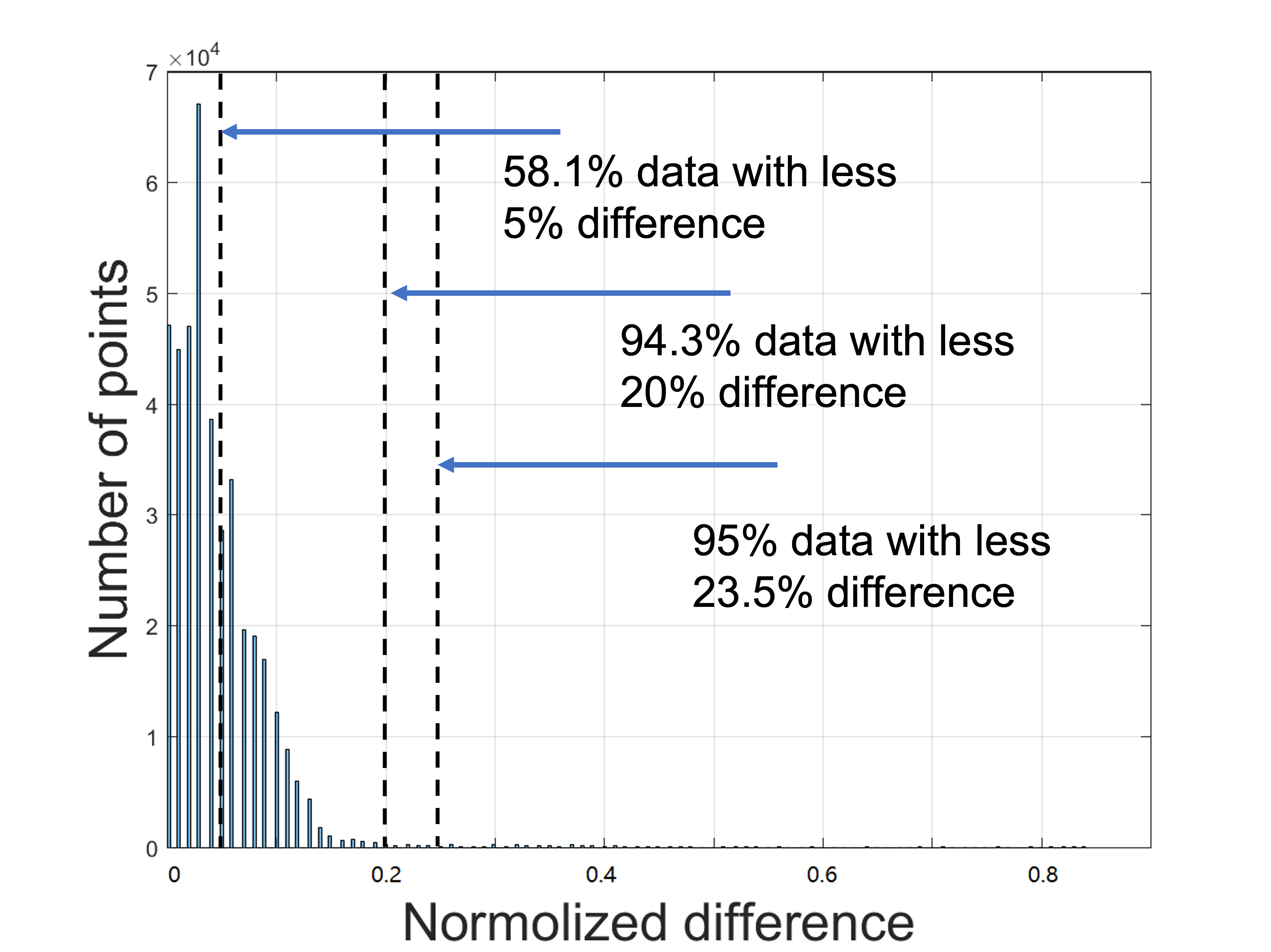}}
\caption{Histogram of normalized prediction difference between WinProp and DCEM in TX Site 2, as shown in Fig.~\ref{fig7}. The normalized difference is calculated by taking the difference between the results from WinProp and DCEM and dividing with the absolute values computed using WinProp. About 58.1\% predictions have less 5\% difference, and about 94.3\% of the data has differences less than 20\%.}
\label{fig9}
\end{figure}
\\The histogram comparison of Site 1 and Site 3 is also attached below.
\begin{figure}[htbp]
\centerline{\includegraphics[scale=0.27]{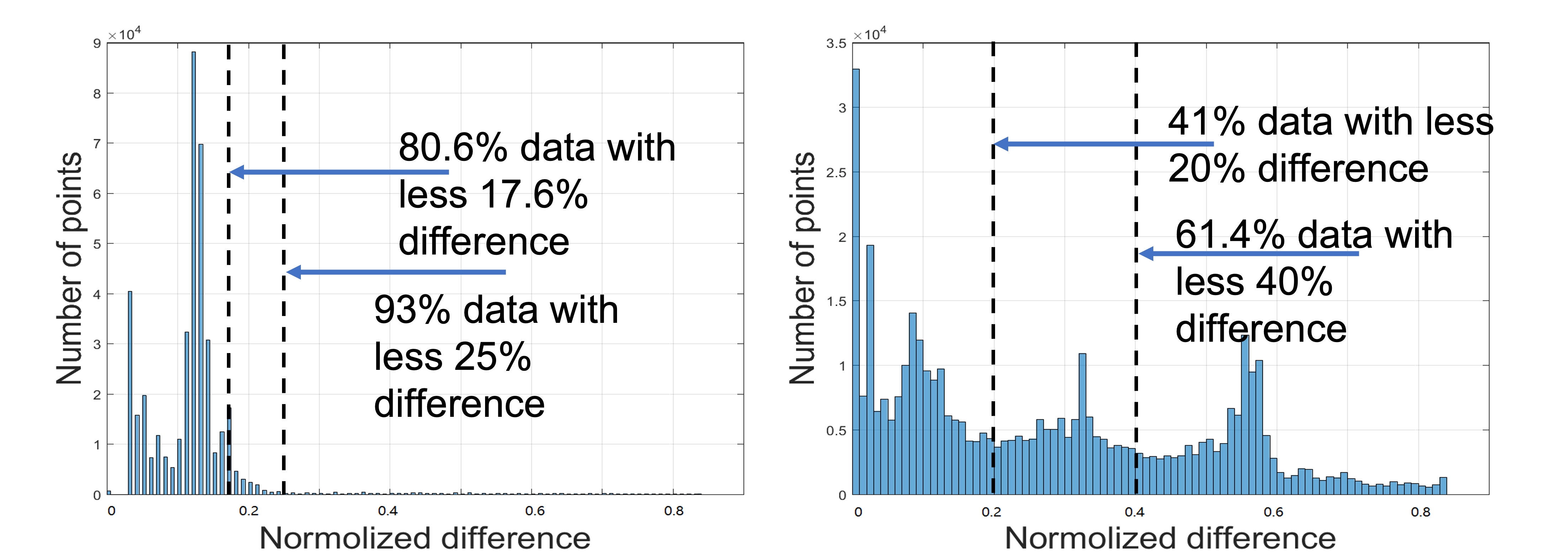}}
\caption{Histogram of normalized prediction difference between WinProp and DCEM in TX Site 1 (left) and Site 3 (right).}
\label{fig10}
\end{figure}

\subsection{Runtime Comparison}\label{AA}

We summarize the working time of the simulations in the following tables.
The parameter settings of the comparison between $GEMV^2$ and DCEM are: $GEMV^2$ runs with reflection and diffraction enabled and DCEM runs with 2000 shooting rays, max 2 reflections, 2 transmissions, and 1 diffraction. We run the tests three times to mitigate computer performance variation. We see from the three test case runs that DCEM has an average of 36.5\% speedup compared to $GEMV^2$, which might results from the both fast ray tracing algorithm improvements and software differences ($GEMV^2$ in MATLAB and DCEM in C++).\\
\begin{table}[htbp]
\caption{Running time COMPARISON of $GEMV^2$ and DCEM}
\begin{center}
\begin{tabular}{|c|c|c|c|}
\hline
\cline{2-4} 
\textbf{Simulator} & \textbf{\textit{Runtime test1}}& \textbf{\textit{Runtime test2}}& \textbf{\textit{Runtime test3}} \\
\hline
$GEMV^2$ & 33.5s& 32.3s& 32.3s \\
\hline
DCEM& 20.5s& 21.2s& 20.6s \\
\hline
\end{tabular}
\label{tab2}
\end{center}
\end{table}

Table IV shows the comparison between Winrprop and DCEM with the whole simulation process, including parsing the modeled village file. WinProp takes a longer time. In terms of the overall preprocessing time, DCEM has improved efficiency in terms of running time by 30.4\%, as compared to WinProp. When taking only the ray tracing simulation time into consideration or performing ray tracing in a simple indoor environment, we set the parameters as follows: WinProp resolution range is 0.02m and it takes max 2 reflections, 2 transmissions, and 1 diffraction by 3D ray tracing model, while DCEM runs with 2000 shooting rays, max 2 reflections, 2 transmissions, and 1 diffraction, Table IV shows that WinProp has a slightly better, but similar computation time to DCEM.

\begin{table}[htbp]
\caption{Running time COMPARISON of WinProp and DCEM}
\begin{center}
\begin{tabular}{|c|c|c|}
\hline
\cline{2-3} 
\textbf{Simulator} & \textbf{\textit{Runtime test1}}& \textbf{\textit{Runtime test2}}\\
\hline
WinProp&  60min& 65min \\
\hline
DCEM& 45min& 42min \\
\hline
\end{tabular}
\label{tab3}
\end{center}
\end{table}

\begin{table}[htbp]
\caption{Running time COMPARISON of WinProp and DCEM excluding environment input processing}
\begin{center}
\begin{tabular}{|c|c|c|c|}
\hline
\cline{2-4} 
\textbf{Simulator} & \textbf{\textit{Runtime case1}}& \textbf{\textit{Runtime case2}}& \textbf{\textit{Runtime case3}} \\
\hline
WinProp& 26s& 25s& 29s \\
\hline
DCEM& 30s& 31s& 35s \\
\hline
\end{tabular}
\label{tab3}
\end{center}
\end{table}

\section{Conclusion}
In this paper, we proposed a reliable and fast RT simulator at EM bands for a typical vehicular communication scenario with a novel dynamic ray tracing algorithm that exploits spatial and temporal coherence. We performed simulations to compare our method with two other well-known RT simulators. The Ericsson's 5G demo from a recent NVIDIA Omniverse keynote speech \cite{b34} showed a ray tracing simulation between base stations and vehicles in a complex environment. Therefore, our next step will be to improve the capabilities of the DCEM to perform RT simulations in more complex and dynamic environments in real time.

\section{Appendix}
We also include LOS and NLOS comparisons between NYUSIM and DCEM here. The following figures show the environment setups and the simulation results from NYUSIM Fig.~\ref{fig11} - Fig.~\ref{fig12} and from DCEM Fig.~\ref{fig13} - Fig.~\ref{fig14}.
\begin{figure}[htbp]
\centerline{\includegraphics[scale=0.3]{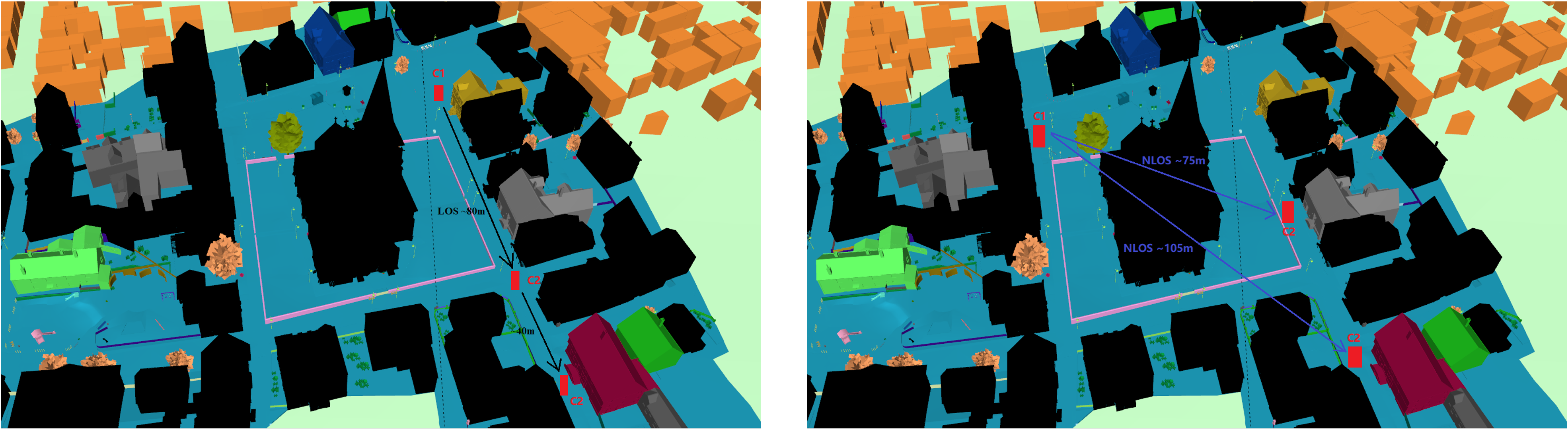}}
\caption{Typical vehicular communication LOS and NLOS links with extended ranges, approximated distances specified in the figures.}
\label{fig10}
\end{figure}
\begin{figure}[htbp]
\centerline{\includegraphics[scale=0.27]{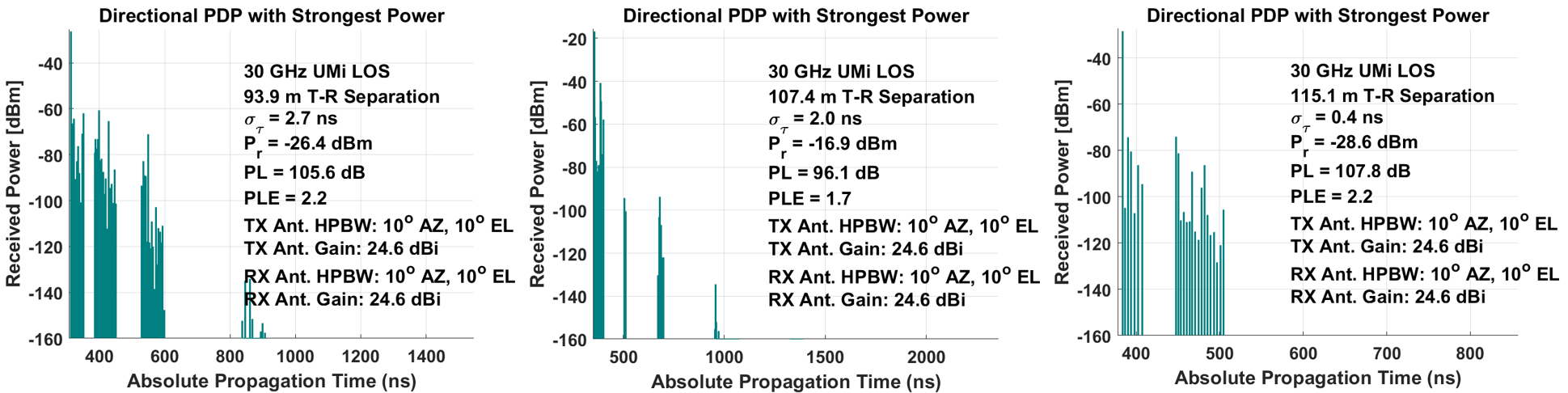}}
\caption{NYUSIM LOS PDP from 93m to 115m, however, since NYUSIM does not take in the environment information, and thus the distance change can not track the moving vehicle.}
\label{fig11}
\end{figure}
\begin{figure}[htbp]
\centerline{\includegraphics[scale=0.27]{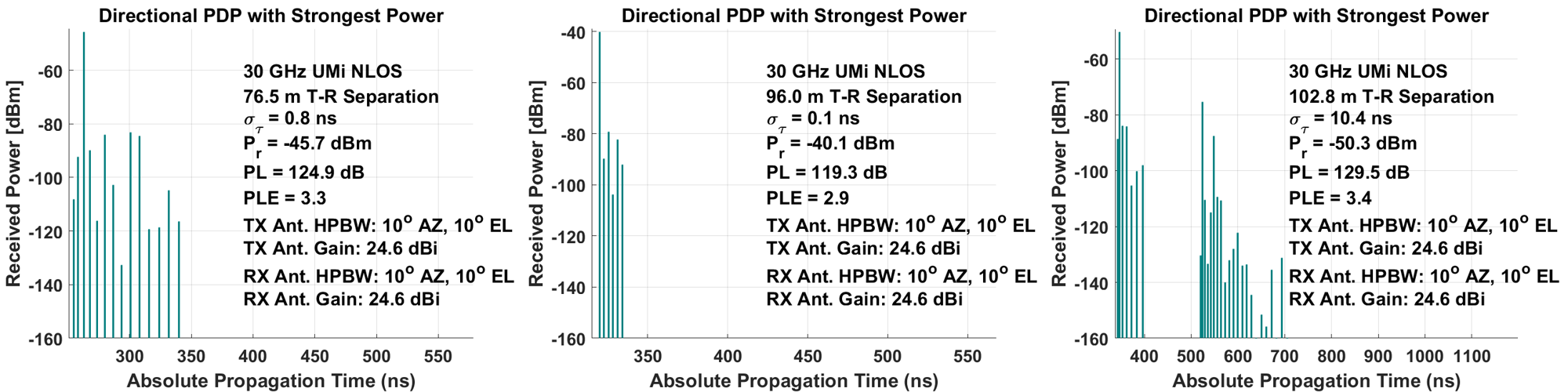}}
\caption{NYUSIM NLOS PDP from 93m to 115m, we see NLOS scenario does experience more path loss, but does not impact the multipath propagation.}
\label{fig12}
\end{figure}
\begin{figure}[htbp]
\centerline{\includegraphics[scale=0.23]{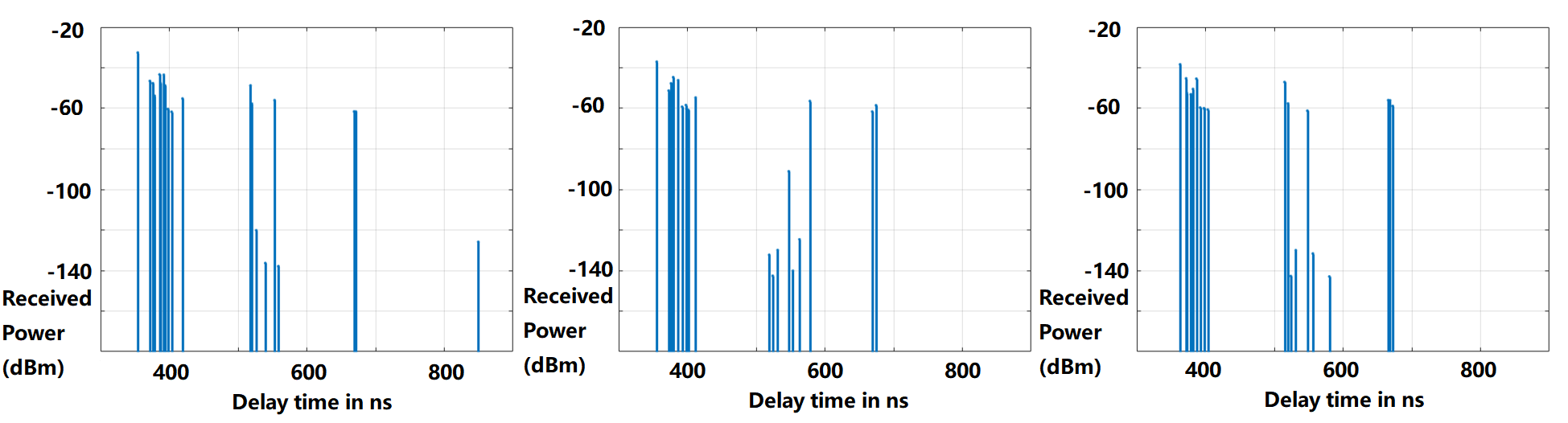}}
\caption{DCEM LOS PDP from 80m to 100m, we see the decreased direct path received power and longer delay as the car moving away, and the clustering of subpaths changing in a consistent manner.}
\label{fig13}
\end{figure}
\begin{figure}[htbp]
\centerline{\includegraphics[scale=0.23]{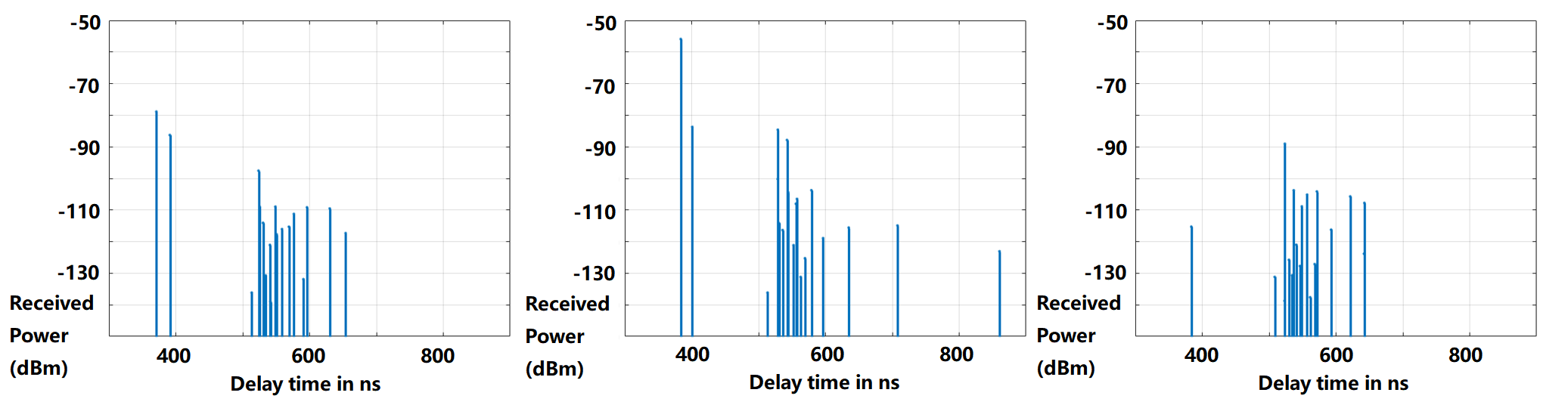}}
\caption{DCEM NLOS PDP from 75m to 105m. We observe that the reduced power and the PDPs are very different at each step according to the environment and changes as the car moves out of sight. However, we observe the main cluster of subpaths at 500-600ns which suggests that the common valid paths are recorded and thus saves the computation time.}
\label{fig14}
\end{figure}
\\These results showed that the proposed algorithm works well with moving object in urban environment. Our future work will be simulating more complex scenes with more moving objects and complex building blocks.

\end{document}